\documentclass[runningheads]{llncs}
\usepackage[T1]{fontenc}
\usepackage{graphicx}
\usepackage{graphicx}

\usepackage{algorithm}
\usepackage{algorithmic}

\usepackage{tabularx} % 
\usepackage{booktabs}
\usepackage{color} % textcolor
\usepackage{float} % 
\usepackage{caption} % 
\usepackage{subcaption} % 

\begin{document}
\title{iLearnRobot: An Interactive Learning-Based Multi-Modal Robot with Continuous Improvement}
\titlerunning{Interactive Learning-Based Multi-Modal Robot}
\author{Kohou Wang$\dag$\inst{1,2} \and
ZhaoXiang Liu$\dag$\inst{*1,2} \and
Lin Bai\inst{3} \and
Kun Fan\inst{3} \and
Xiang Liu\inst{1,2} \and
Huan Hu\inst{1,2} \and
Kai Wang\inst{1,2} \and
Shiguo Lian\inst{*1,2}}
\authorrunning{K. Wang et al.}
\institute{Unicom Data Intelligence, China Unicom \and
Data Science \& Artificial Intelligence Research Institute, China Unicom\and
China United Network Communications Group Corporation Limited \\
\email{\{wangzp103, liuzx178, bail16, fankun2, liux750, huh30, wangk115, liansg\}@chinaunicom.cn} \\
$\dag$Equal contribution,*Corresponding author(s)
}
\maketitle              % typeset the header of the contribution
\begin{abstract}
    It is crucial that robots' performance can be improved after deployment, as they are inherently likely to encounter novel scenarios never seen before. This paper presents an innovative solution: an interactive learning-based robot system powered by a Multi-modal Large Language Model(MLLM). A key feature of our system is its ability to learn from natural dialogues with non-expert users. We also propose \textit{chain of question} to clarify the exact intent of the question before providing an answer and \textit{dual-modality retrieval} modules to leverage these interaction events to avoid repeating same mistakes, ensuring a seamless user experience before model updates, which is in contrast to current mainstream MLLM-based robotic systems. Our system marks a novel approach in robotics by integrating interactive learning, paving the way for superior adaptability and performance in diverse environments. We demonstrate the effectiveness and improvement of our method through experiments, both quantitively and qualitatively.
\end{abstract}

\section{Introduction}
The development of multimodal robots has been a highly regarded field, especially in 
recent years, the rapid development of Large Language Models(LLMs)~\cite{llm-survey} and Multimodal Large Language Models(MLLMs)~\cite{mllms} has greatly enhanced the capabilities 
of robots~\cite{driess2023palm,multimodal-robot1,MuModaR}. 
The demand for continuously 
    promoting robots' capabilities to adapt to specific domains and novel scenarios after deployment grows ever more intense.
    However, existing works mainly focus on adapting robots' manipulation and navigation
    to specific domains and novel scenarios
    ~\cite{vison-language-navigation,justAsk,distillingRobot,mqa}, while few 
    pay attention to visual perception, understanding and recognition.

A typical way to optimize performance of visual perception, understanding and recognition is to assemble a team of domain experts to collect, clean, and annotate data in specific scenarios, which is then used to train and update the model's parameters. However, given the vast scenarios in which robots are applied, this method is clearly not economically viable. 

Beyond that, two issues consistently hang over all systems that interact with humans: 
\begin{itemize}
    \item In daily conversations, non-expert users' questions are often ambiguous, making it difficult to discern their exact intent. However, MLLMs are typically constrained to provide responses to these equivocal queries, which greatly impacts user interaction experience. 
    \item Inevitably, MLLM-based robot sometimes provides incorrect answers, a problem that is exacerbated when encountering scenarios they have not seen before. Yet, improving these responses requires waiting for substantial resources and time (the time cost being particularly intolerable to users) to update the MLLMs before correct answers can be given.
\end{itemize}
Here come questions: Should we simply allow the robots to keep repeating the same mistakes before the model is updated? What should be done to ensure a better user experience?

\begin{figure}[t!]
    \centering
    \includegraphics[width=1\linewidth]{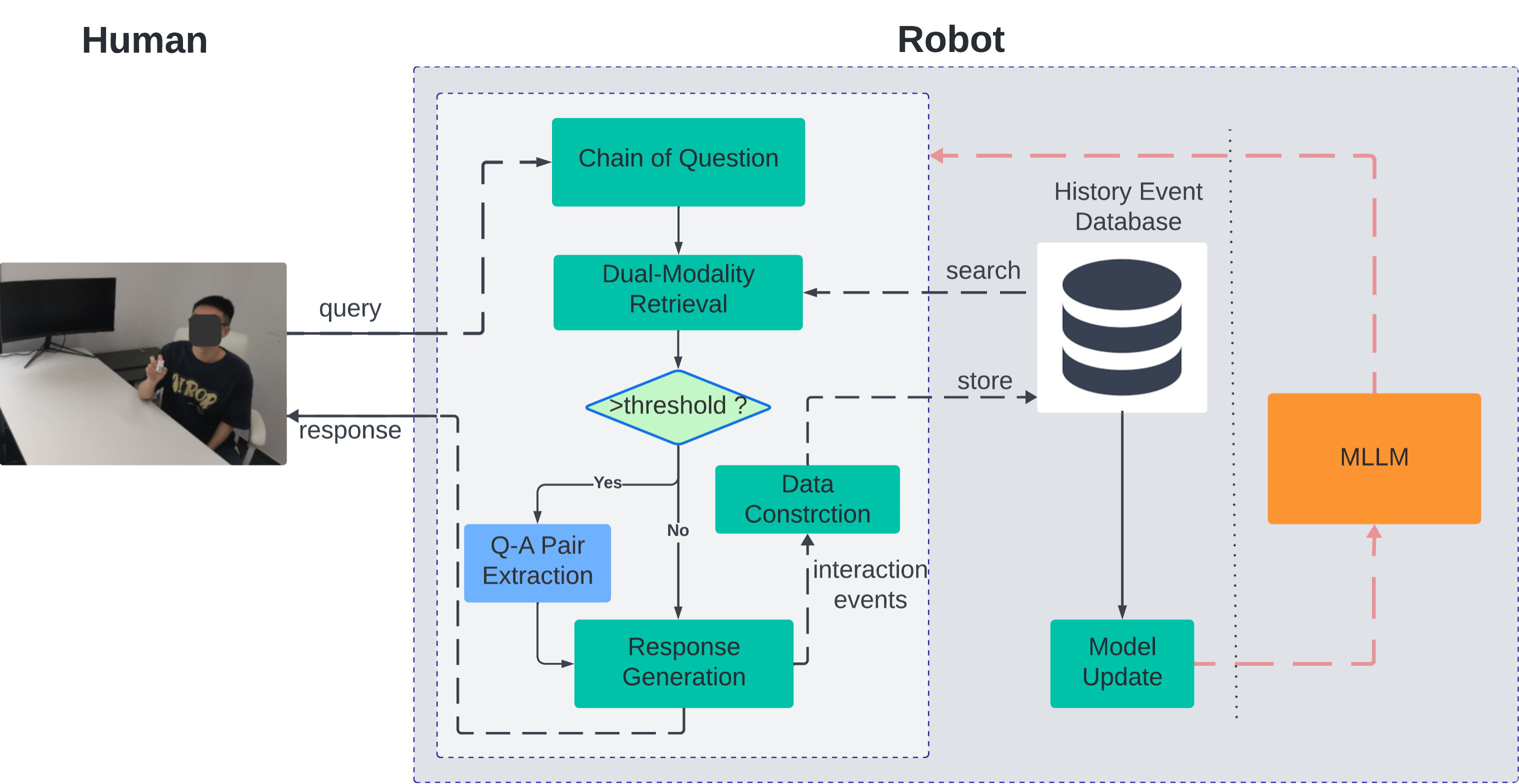} 
    \caption{Flowchart of iLearnRobot. The robot will first attempt to understand the user's true intent through multiple rounds of questioning, it then checks whether there are similar event in the history event database to base its response on. Ultimately, this round of dialogue will also be recorded in the history event database and used to update the MLLM.
    }
    \label{flowchart-v3}
    \end{figure}

To address these issues, this paper presents an 
    innovative framework: 1) to pose questions before answer a question; 2) to learn from natural dialogues with non-expert users, and also,
    3) to retrieve history interaction events to avoid repeating same mistakes before model updates.

In daily interactions between robots and humans, verbal corrections made by humans can be leveraged. 
In our framework, the interactive learning-based robot, powered by MLLM, 
utilizes our designed chain of question module to iteratively ask the user additional questions
to elicit more specific details about their query in everyday dialogues, until it can confidently ascertain the user's precise intent. 
Then our dual-modality retrieval module will retrieve past similar event to serve as a reference for the final response.
Subsequently, the robot, utilizing response generation module, will answer the question conditioned on past interaction event, 
user's precise query and image of the scenario. These interaction events will then be distilled into specific formats and stored in the history event database utilizing data construction module, which will be used for retrieval and updating models.

The contributions are:

\begin{itemize}
\item We propose a novel framework by integrating interactive learning into multi-modal robotics, allowing the robot to learn from daily natural dialogues and continuously improve its performance;
\item We propose \textbf{chain of question} module that clarifies the exact intent of users' questions before providing answers, ensuring a better user experience ; 
\item We propose \textbf{dual-modality retrieval} module that ensures the robot generates more accurate responses based on past interaction events before the updating of model, to avoid repeating the same mistakes corrected before, leading to better adaptability in diverse environments; 
\item We evaluate our framework and method both quantitively and qualitatively, and the results demonstrate their effectiveness.
\end{itemize}

\section{Related Work}
Before MLLMs was applied to the field of robotics, there were also some efforts in the Natural Language Processing(NLP)~\cite{kang2020natural} field to enhance their capabilities.
MM-REACT~\cite{yang2023mm} empowers ChatGPT with visual understanding capabilities, focusing on enhancing ChatGPT's visual understanding capabilities and utilizes prompting and reasoning/action texts to interact with vision experts, 
Art~\cite{paranjape2023art} leverages a frozen LLM to automatically generate multi-step reasoning and utilize external tools for complex tasks. These methods focus on using external tools to enhance certain aspects of the model's capabilities without optimizing the model's adaptability and without updating the model to accommodate new scenarios.
In contrast, we propose an innovative MLLM-based robot system that focuses on interactive learning and continuous improvement through natural dialogue with non-expert users, leveraging the MLLM's own capabilities, employing our proposed chain of question, dual-modality retrieval, and model updates modules.

\begin{figure}[t!]
    \centering
    \includegraphics[width=0.6\linewidth]{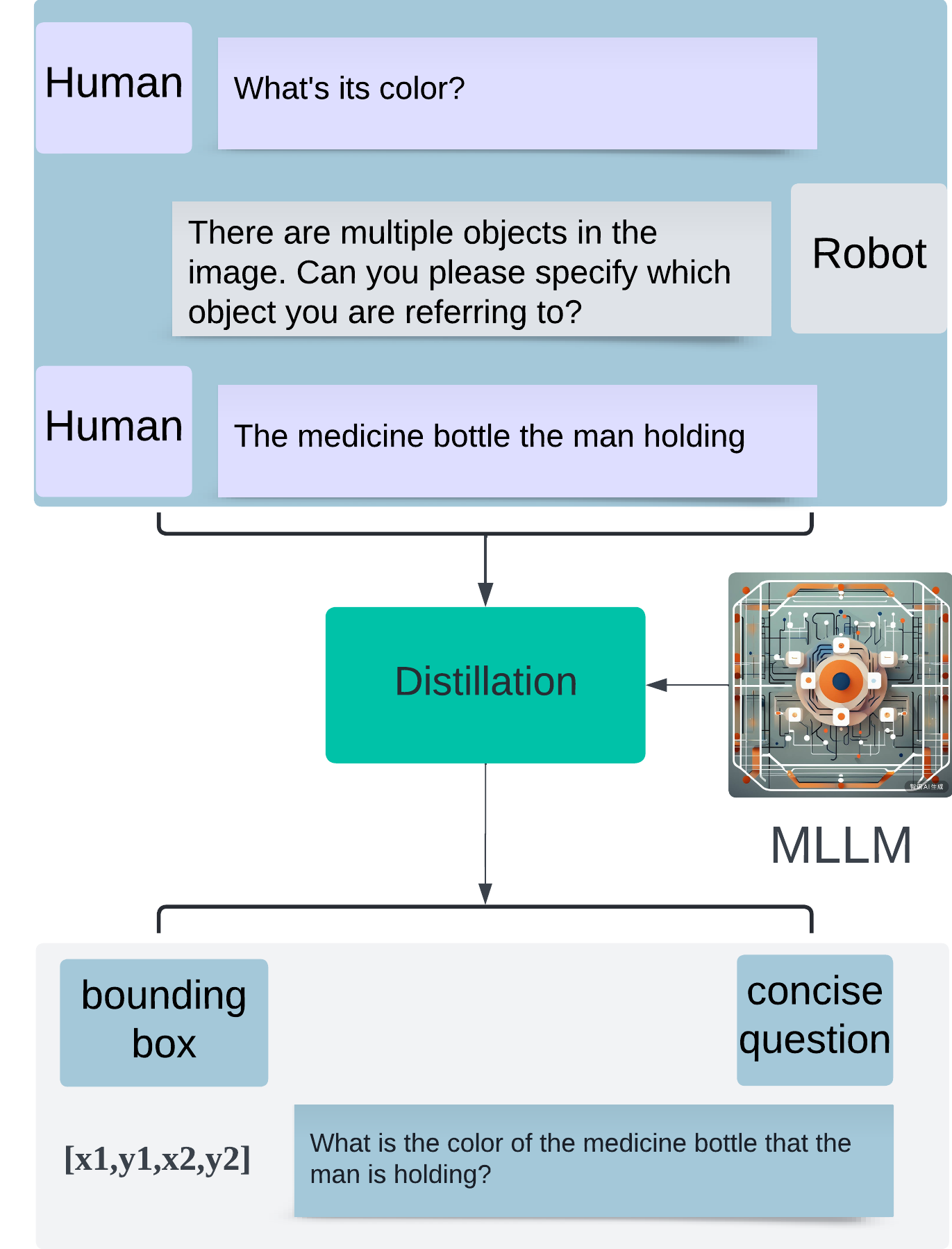} % 
    \caption{Illustration of our \textit{chain of question} module. 
    }
    \label{chain-of-question}
\end{figure}
The integration of MLLMs into robotics has been a focal point. These approaches 
strive to empower robots to process and understand various types of data, and primarily focus on the domains of navigation and manipulation. 

In the field of navigation, Chi et al~\cite{justAsk} proposes an interactive learning framework to endow the agent with the ability to ask
for users' help, which is among the first to introduce human-agent interaction in the instruction-based navigation 
task. Zhu et al~\cite{vison-language-navigation} introduce a framework with four self-supervised auxiliary 
reasoning tasks to take advantage of the additional training signals derived from the semantic information, which 
helps the agent to reason about its activity. 

In the domain of manipulation, Deng et al~\cite{mqa} propose a framework that integrates a Question Answering module and a manipulation 
module to accomplish a newly defined task, where the robot performs manipulation actions to change the environment 
in order to answer a given question. Kenfack et al~\cite{robotvqa} propose an architecture for robot vision on manipulation
tasks, which encompasses a generator of training datasets and a learnable scene describer for Robot Visual Question Answering. 
Zha et al~\cite{distillingRobot} present a large language model-based system that can respond to arbitrary forms of language feedback, and distill generalizable knowledge from corrections. Thomason et al~\cite{Improving-grounded} present an end-to-end pipeline
for translating natural language commands to discrete robot actions, and use dialogs to jointly improve language parsing and concept grounding.

However,
the broader challenges of adapting to novel scenarios and updating the MLLM's intrinsic capability of perception, understanding and recognition are less addressed. This paper aims to bridge this gap by introducing a novel robot system powered by MLLM, leveraging interactive learning and dual-modality retrieval to update model's capabilities and guarantee a decent user experience before the updating of model.

\section{Methods}
The flowchart of our framework is shown in figure~\ref{flowchart-v3}.
During the natural dialogues with users, the robot will first attempt to thoroughly
understand the user's true intent through multiple rounds of questioning, 
it then checks whether there are similar event in the history event database to base its response on. Ultimately, this round of dialogue will also be recorded in the history event database and used to train and update the MLLM.

\subsection{User Interaction}

The system utilizes LLaVA-NeXT~\cite{llava-next},
a multi-modal large language model, 
to analyze the context and generate a verbal response. LLaVA-NeXT, with improved reasoning, OCR, and world knowledge, its ability to
understand and generate natural 
language is crucial in ensuring that the robot's responses are both coherent and accurate.

\begin{figure}[t!] 
    \centering
    \includegraphics[width=0.7\linewidth]{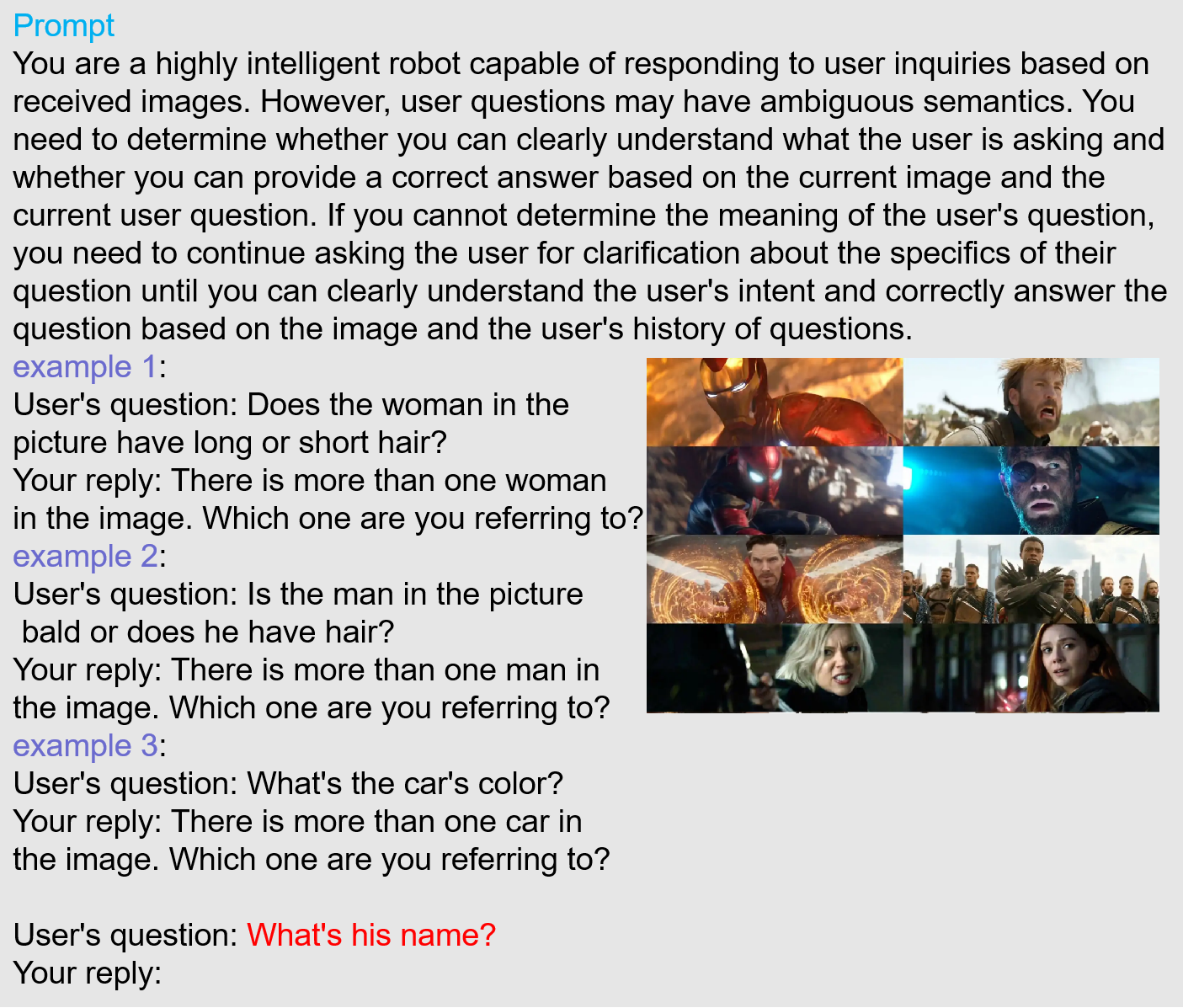}
    \caption{Prompt used for chain of question. The subsequent dialogue for this quesiton can be seen in figure~\ref{prompt-distilling-conversation}.
    }
    \label{question-loop-prompt}
    \end{figure}

\begin{figure}[t!] 
    \centering
    \includegraphics[width=0.7\linewidth]{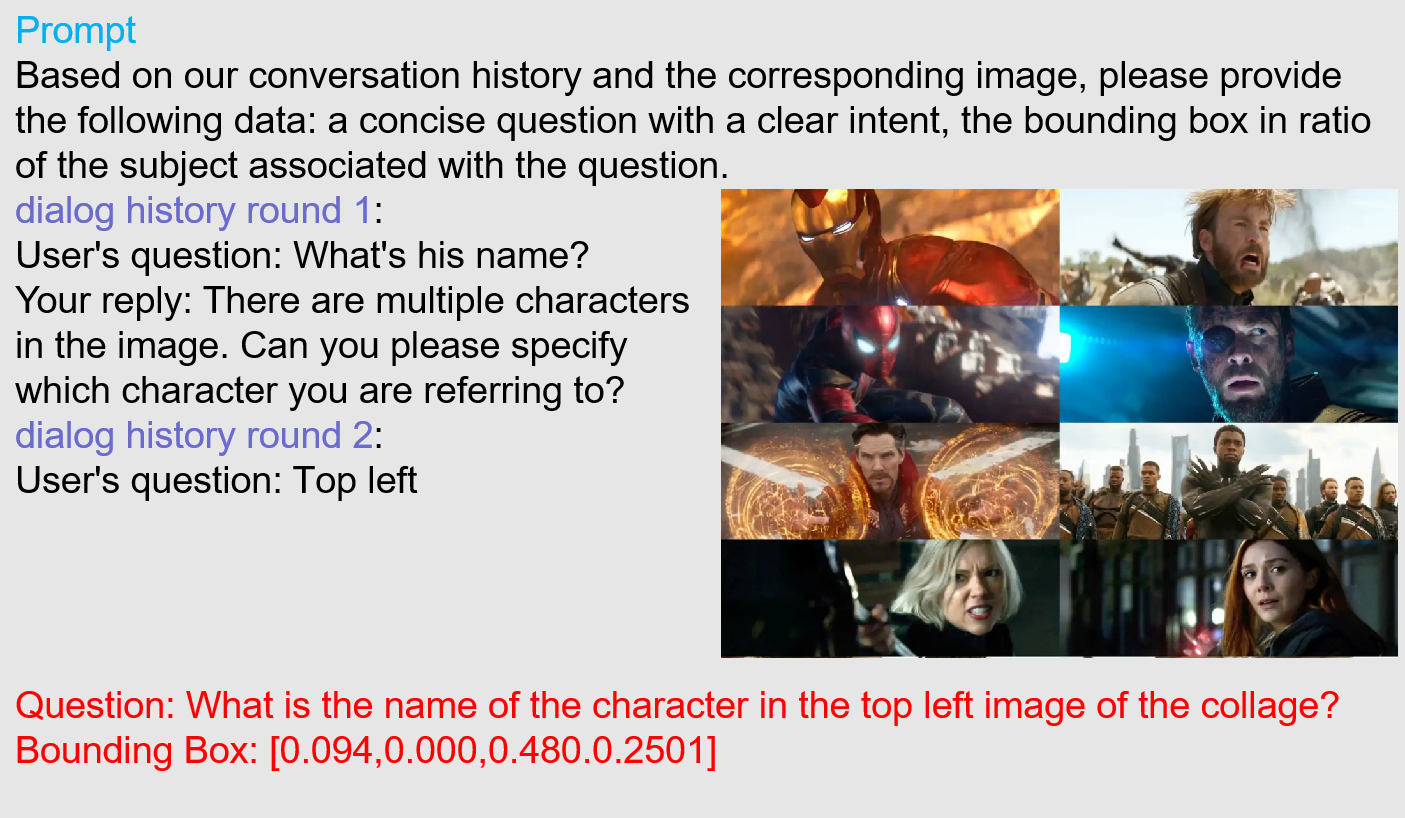}
        \caption{Prompt used for distilling data. These user natural dialogues are distilled into concise data. %Sentences highlighted against a yellow background denote scenarios in which an answer is available.
        \textcolor{red}{Red sentences} are distilled results.}
        \label{prompt-distilling-conversation}
    \end{figure}

% \begin{figure}[t!]
%     \centering
%     \begin{subfigure}[b]{0.45748\linewidth}
%     \includegraphics[width=\linewidth]{images/question-loop-prompt.png}
%     \caption{Prompt used for chain of question. The subsequent dialogue for this quesiton can be seen in figure~\ref{prompt-distilling-conversation}.
%     }
%     \label{question-loop-prompt}
%     \end{subfigure}
%     \hfill 
%     \begin{subfigure}[b]{0.45748\linewidth}
%         \includegraphics[width=\linewidth]{images/prompt-distilling-conversation-v3.png}
%         \caption{Prompt used for distilling data. These user natural dialogues are distilled into concise data. %Sentences highlighted against a yellow background denote scenarios in which an answer is available.
%         \textcolor{red}{Red sentences} are distilled results.}
%         \label{prompt-distilling-conversation}
%     \end{subfigure}
%     \caption{Prompts used for a): chain of question and b): distilling data. }
%     \label{fig:formulation}
% \end{figure}

\textbf{Chain of Question.} Imagine a scenario: your robotic system is highly intelligent, yet a user approaches and asks a vague question: "What is that?" Despite the robot's capabilities, it's as if a scholar has met a soldier, unable to put its skills to use. Moreover, MLLMs are typically constrained to provide responses to these equivocal questions, so your robot is forced to make assumptions about what the user is referring to and offer an answer. If this is not what the user meant, it greatly impacts the user's interaction experience.

We propose the \textit{chain of question} module to address this problem. The illustration is shown in figure~\ref{chain-of-question}. This module's function is to clarify the user's precise intent. If the robot is unable to determine the user's exact question based on the current image and query, it initiates a clarification process. The robot poses additional questions to the user, seeking to elicit more specific details about their query. This iterative process continues until the robot can confidently ascertain the user's intent. The prompt is shown in figure~\ref{question-loop-prompt}.
During the clarification process, the robot collects additional information about the user's query 
through several rounds of supplementary questioning, until it determines during the conversation
phase that it has fully understood the user's precise intent. This interactive process allows the robot to improve 
    its understanding of user queries, 
    leading to more accurate and effective dialogues.

    Then, given previous rounds of dialogues and the corresponding image, the robot is prompted to generate the coordinate location 
of the subject related to the question within the image, represented by a bounding box in ratio form. The specific prompt used for distilling data is shown in figure~\ref*{prompt-distilling-conversation}.
    The robot then extracts the target subject from the image and encodes it into embedding form using the CLIP~\cite{clip} image encoder. Similarly, the robot is prompted to distill from the previous dialogues a concise question that encapsulates the user's precise intent. This question is also encoded into an embedding form using the CLIP text encoder.% 
\begin{figure}[t!] 
    \centering
    \includegraphics[width=0.5\linewidth]{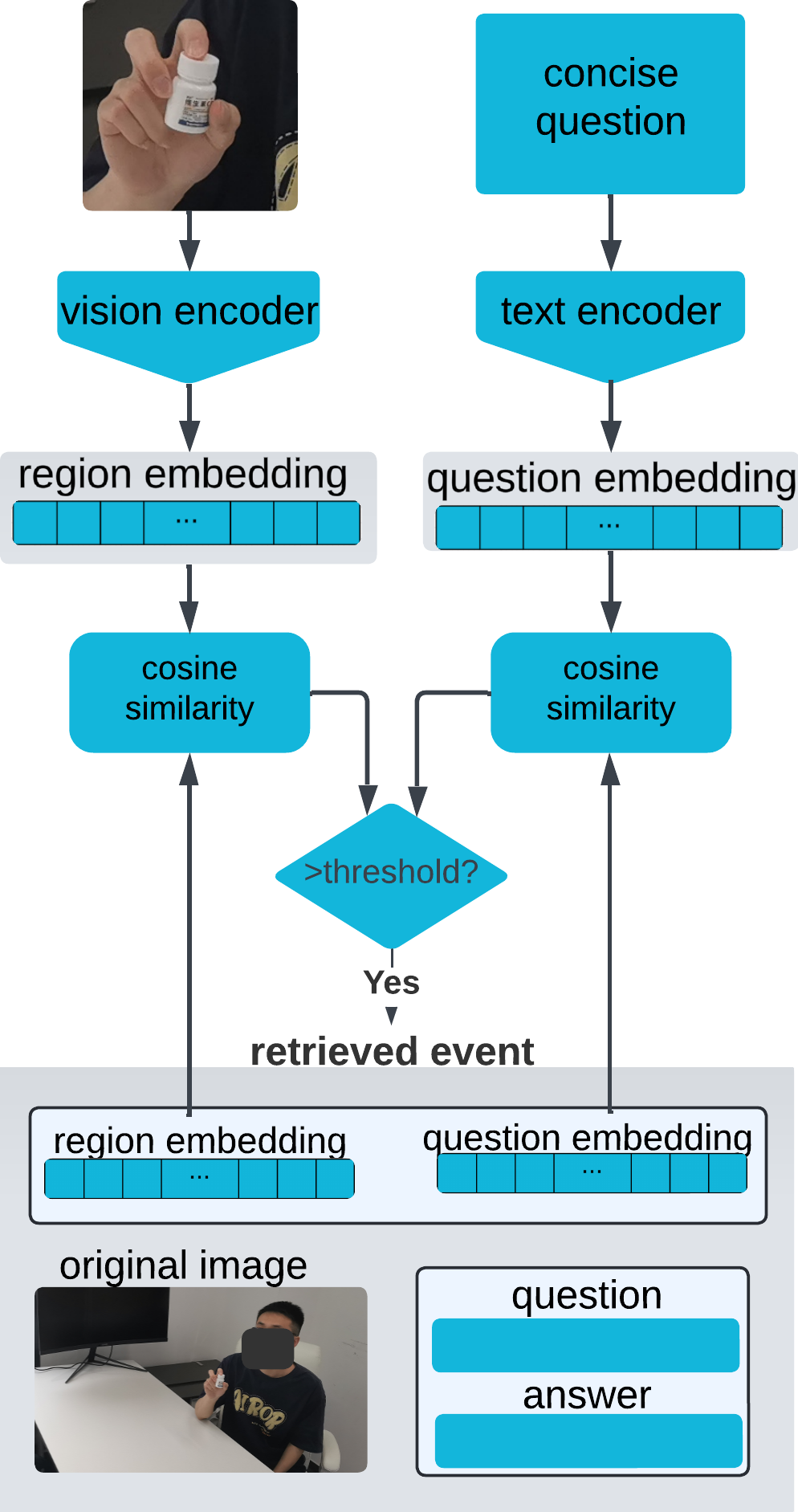} 
    \caption{Illustration of our \textit{dual-modality retrieval} module.
    }
    \label{dual-modality-retrieval}
    \end{figure}

\textbf{Dual-Modality Retrieval.} 
In real-world applications, robotic systems will inevitably make various mistakes. However, correcting these errors is often time-consuming and labor-intensive, which can be particularly intolerable to users in terms of time. Before the model is updated, the robotic system will repeatedly make the same mistakes. Inspired by~\cite{distillingRobot}'s approach that can distill generalizable knowledge from corrections in the domain of manipulation, we propose the \textit{dual-modality retrieval} module.

This module becomes essential for improving the overall interaction quality and ensuring users' satisfaction. 
The illustration is shown in figure~\ref{dual-modality-retrieval}. 

Specifically, given a data sample $(I_i, R_i, Q_i)$ containing the original image $I_i$, subject region $R_i$, and user's question $Q_i$, we use the CLIP image encoder $\Phi_{img}$ to extract the image embedding $e_{img}\in R^d$ of $R_i$, and the
CLIP text encoder $\Phi_{text}$ to extract the text embedding $e_{text}\in R^d$ of $Q_i$. The symbol $d$ refers to the feature dimension (e.g., $d = 576$ for CLIP ViT-B/16). 
These embeddings are then used to calculate the cosine similarity with 1) embeddings of the subject regions and 2) the embeddings of the full questions of past interaction events from the historical event database. The past interaction event is in format of $(e_{img}, e_{text}, I_i, Q_i, A_i)$, indicating embeddings, the original image $I_i$, user's question $Q_i$ and correct answer $A_i$. If the cosine similarities each meet a certain specific threshold, it is considered a successful retrieval of a past event from the database. The correct answers included in this past event will then serve as a reference for the current question.The establishment of the history event database will be described in section \textbf{Data Construction}.

\begin{figure}[t!]
    \centering
    \includegraphics[width=0.65\linewidth]{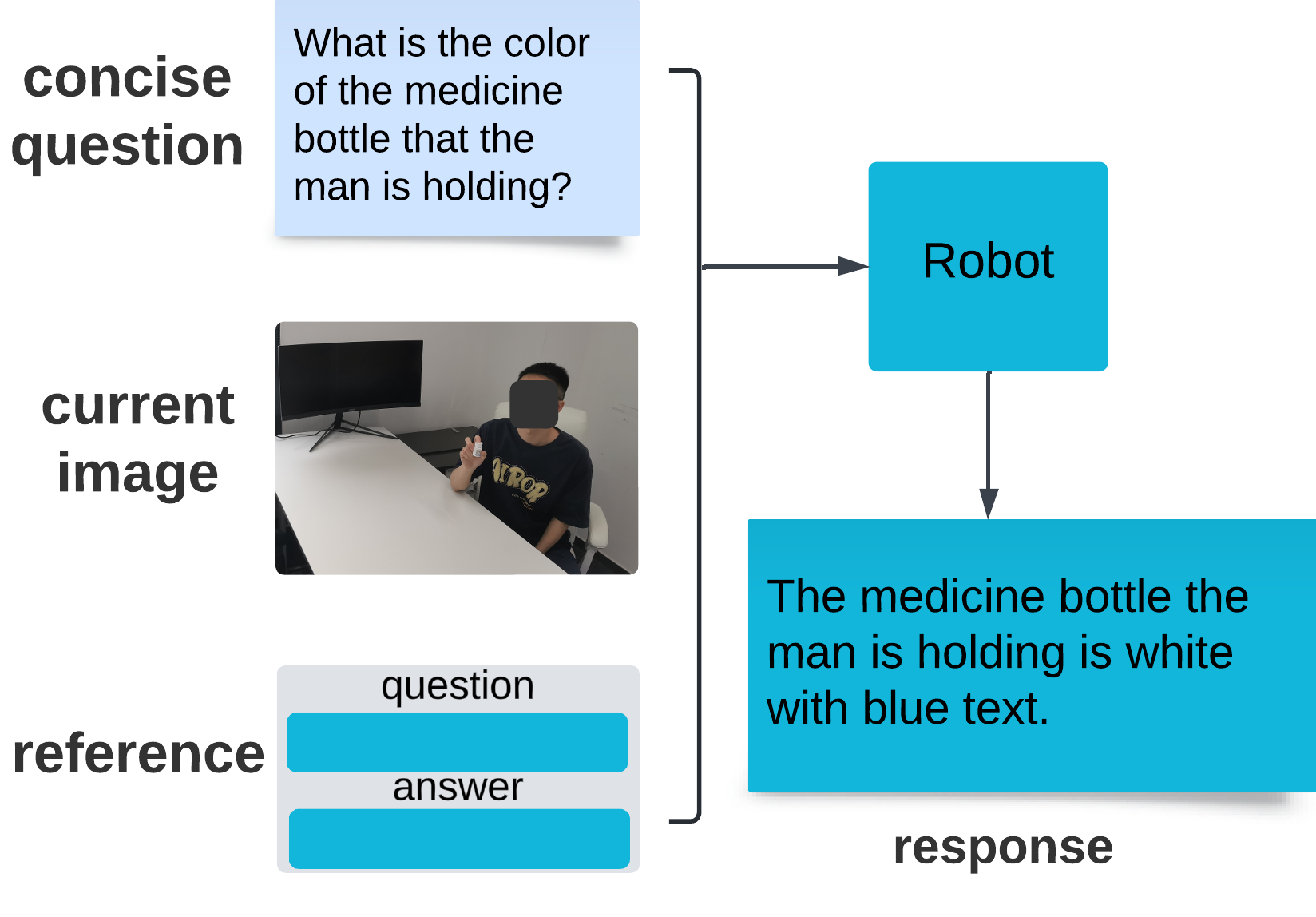} % 
    \caption{Illustration of our \textit{response generation} module.
    }
    \label{response generation}
\end{figure}

\begin{figure}[t!]
    \centering
    \begin{subfigure}[b]{0.7\linewidth}
    \includegraphics[width=\linewidth]{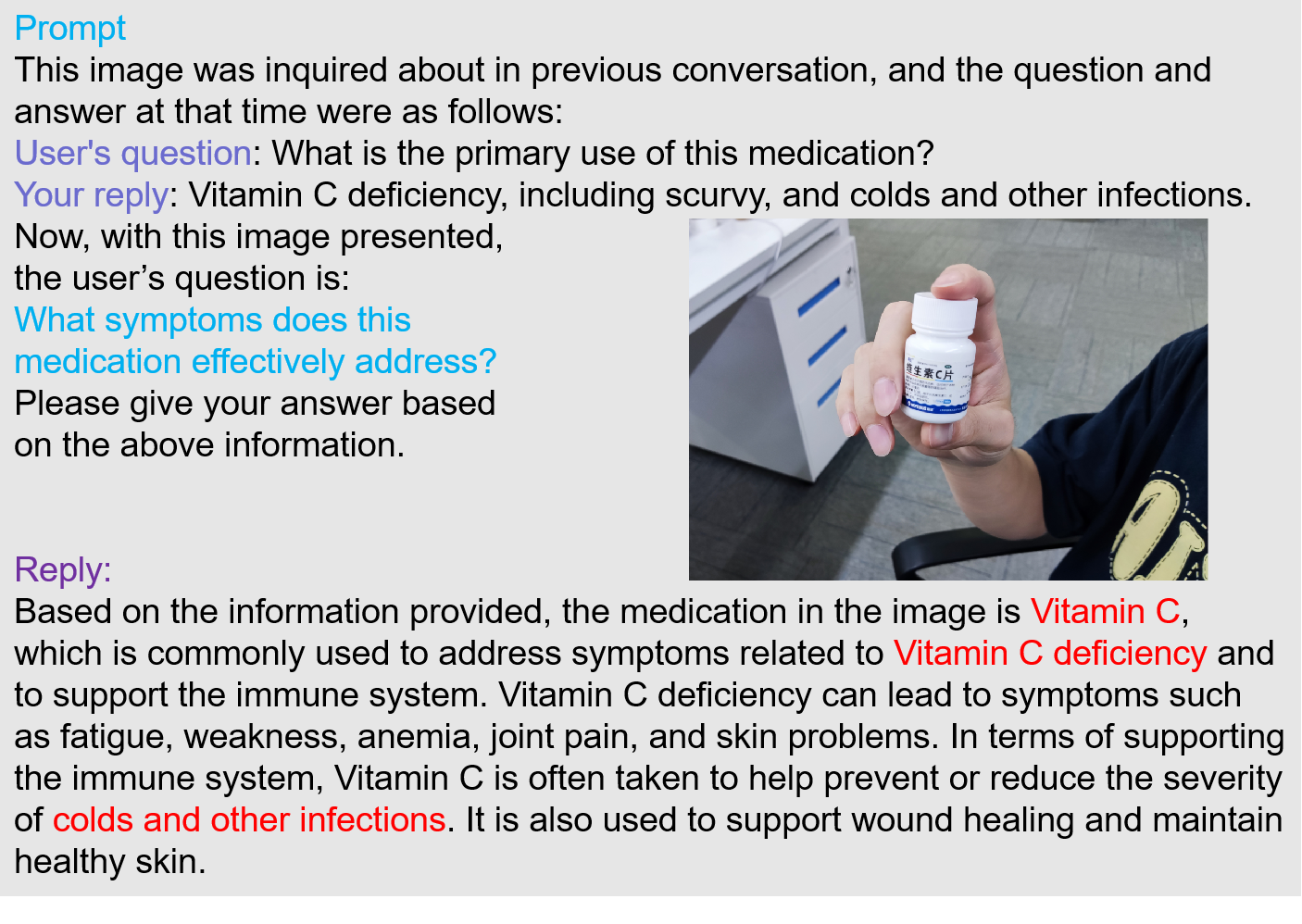}
    % \caption{Prompt used for chain of question. The subsequent dialogue for this quesiton can be seen in figure~\ref{prompt-distilling}.
    % }
    \label{prompt-historyEvent-compare-a}
    \end{subfigure}
    \hfill 
    \begin{subfigure}[b]{0.7\linewidth}
        \includegraphics[width=\linewidth]{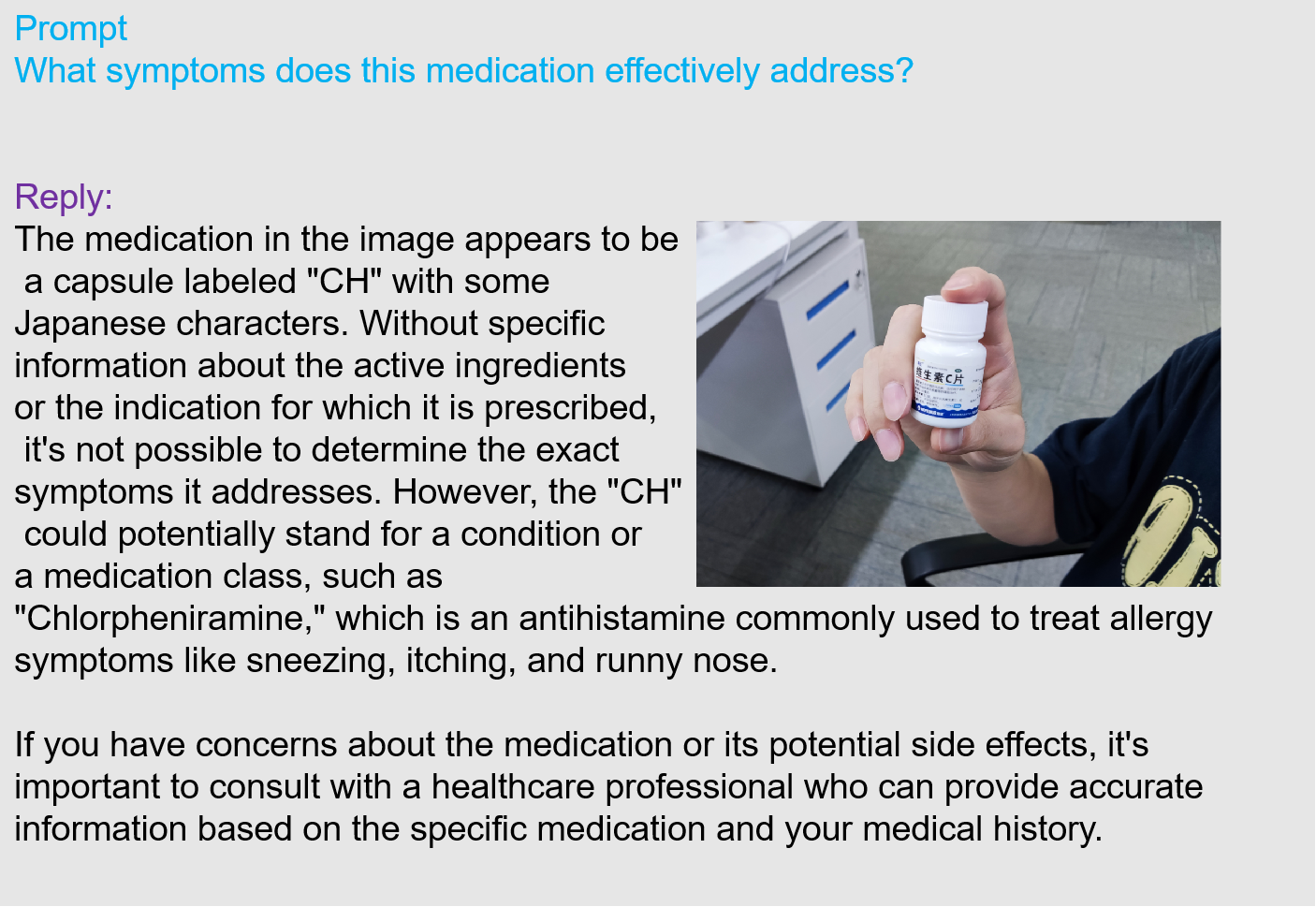}
        % \caption{Prompt used for distilling data. These user natural dialogues are distilled into concise data. %Sentences highlighted against a yellow background denote scenarios in which an answer is available.
        % \textcolor{red}{Red sentences} are distilled results.}
        \label{prompt-historyEvent-compare-b}
    \end{subfigure}
    \caption{Left: prompt and the generated responses. Right: responses generated without this prompt.
    The robot manages to generate correct response based on the retrieved interaction event, 
    while it replies nonsense without the reference.}
    \label{prompt-historyEvent}

\end{figure}

\textbf{Response Generation.} This module is designed to generate final response based on, if available, 
similar past interaction event retrieved from the history event database. The illustration is shown in figure~\ref{response generation}. 
If the dual-modality retrieval stage fails to find a match in the history event database, 
the robot will directly generate responses based solely on the several rounds of dialogs and the input image using its intrinsic knowledge.
However, if the dual-modality retrieval succeeds in retrieving the related history event, 
the system will retrieve the question and answer from the database record and
use them as a reference to generate the final response. The specific prompt and responses generated with/out this prompt are shown in figure~\ref{prompt-historyEvent}. 

\subsection{Data Construction}\label{Data-Construction}

\begin{figure}[t!]
    \centering
    \includegraphics[width=1\linewidth]{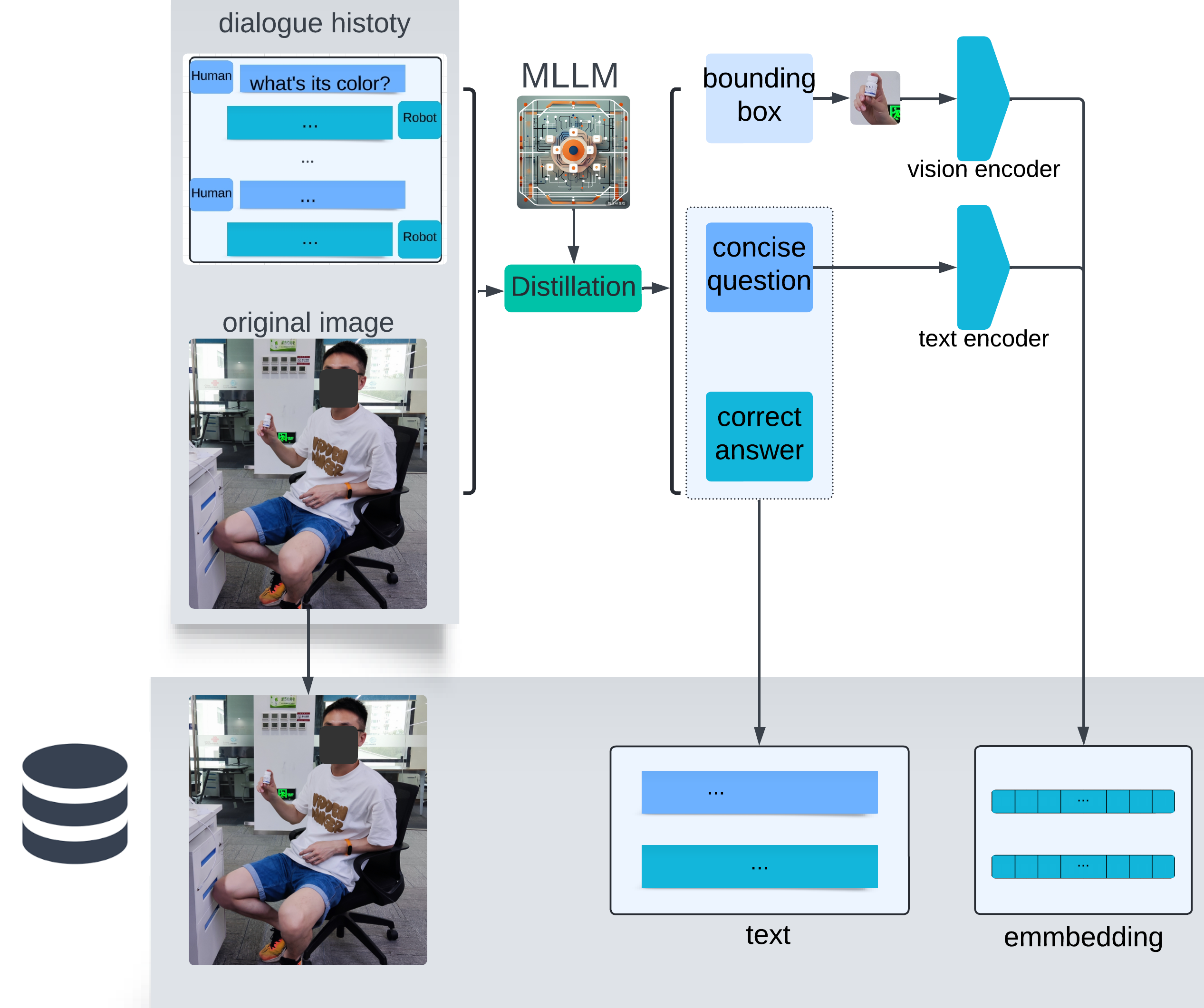} % 
    \caption{Illustration of our \textit{data construction} module.
    }
    \label{data-construction}
\end{figure}

In daily interactions between robots and humans, verbal corrections offered by humans are invaluable, serving as a crucial and exploitable resource for improving robots' performance. We distill such interaction dialogues into a specific format, subsequently storing them in history event database. The illustration is shown in figure~\ref{data-construction}. This module processes complete user interactions, which include multi-round dialogues between 
user and robot, as well as the image associated with the user's query. 

\begin{figure}[t!]
    \centering
    \includegraphics[width=0.748\linewidth]{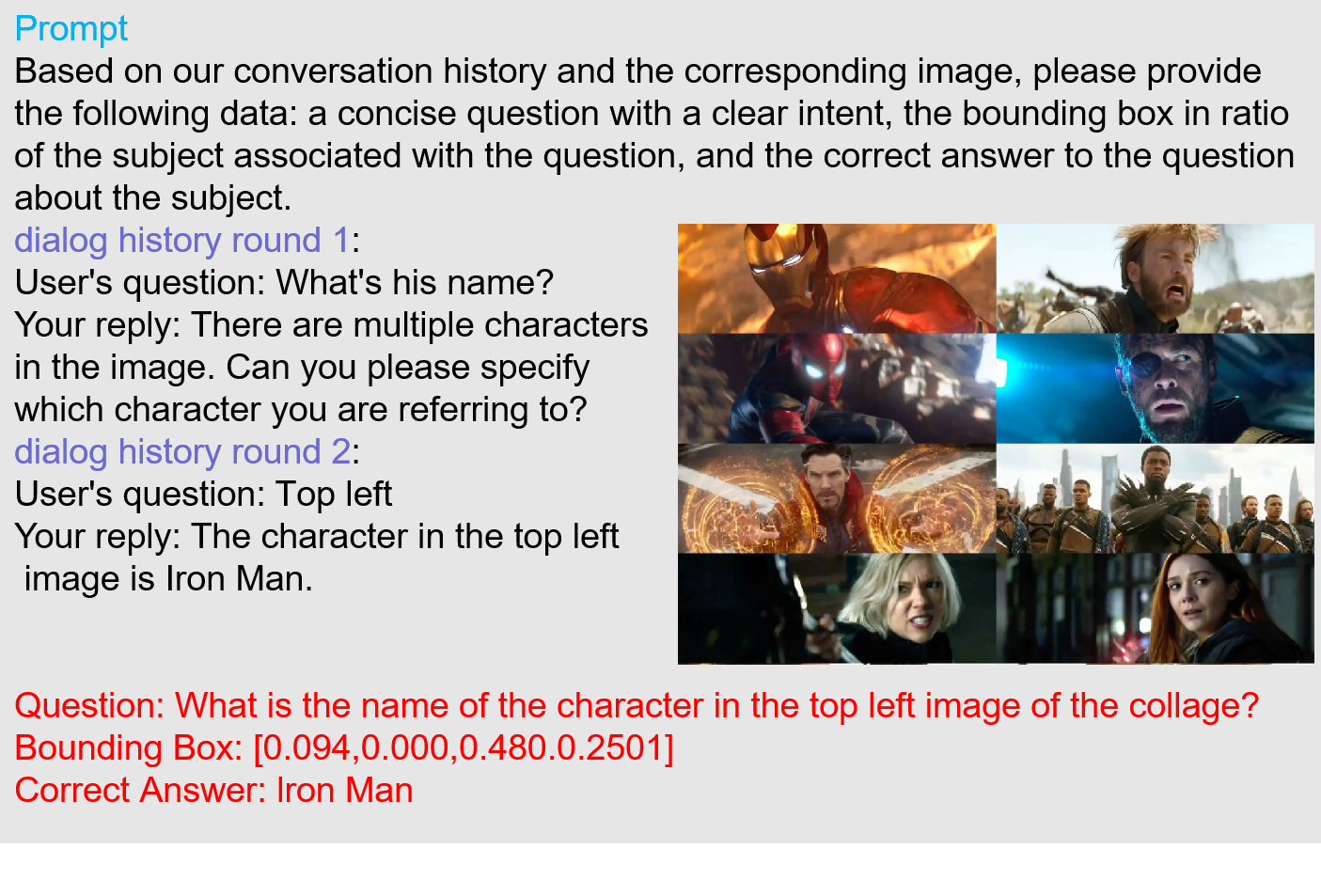} % 
    \caption{Prompt used for distilling natural dialogs into a concise data format that includes \textit{question with clear intent}, \textit{subject bounding-box} and \textit{correct answer}.
    }
    \label{prompt-summary-distill}
\end{figure}

Figure~\ref{prompt-summary-distill} demonstrates the process of distilling these natural dialogues into a concise data format that includes:
\begin{itemize}
\item \textbf{Question with Clear Intent}: The system summarizes the user's query into a single, consise question with a clear intent.
This question captures the essence of the user's inquiry, providing a clear and concise representation of the user's intent;
\item \textbf{Subject Bounding-Box}: The system identifies the bounding box of the subject within the image associated 
with the user's query.
The bounding box allows the system to accurately crop the image to focus on the relevant area of interest;
\item \textbf{Correct Answer}: The system identifies the correct answer to the user's query about the subject. The answer can
be of any property, such as breed, shape, texture, etc.
This answer ensures that the robot can correctly identify and respond to the user's query in later dialogues when questioned
about the same subject's same property.
\end{itemize}

With this distilled data, the system proceeds to extract the target subject region from the image based on 
the generated bounding box. The subject region and user's precise question are then encoded into embeddings,
using the text encoder and image encoder of CLIP.

Specifically, given a data sample $(I_i, R_i, Q_i, A_i)$ containing the original image $I_i$, subject region $R_i$, user's question $Q_i$ and correct answer $A_i$, we use the CLIP image encoder $\Phi_{img}$ to extract the image embedding $e_{img}\in R^d$ of $R_i$, and the
CLIP text encoder $\Phi_{text}$ to extract the text embedding $e_{text}\in R^d$ of $Q_i$, while original image $I_i$, user's question $Q_i$ and correct answer $A_i$ remain unchanged. These embeddings, along with the original image $I_i$, user's question $Q_i$ and correct answer $A_i$, in format of $(e_{img}, e_{text}, I_i, Q_i, A_i)$, are stored in a history event database $H$, where they can be retrieved during user interactions to help the robot to generate more accurate responses and provide better user experiences.

\subsection{Model Update}
Model update initiates when the number of accumulated interaction events 
in history event database $H$ reaches a predefined threshold. The training data, in the format of $(I_i, Q_i, A_i)$, is then fed into the LLaVA-NeXT model, which is fine-tuned one epoch from LLaVA checkpoints with LoRA~\cite{lora} following the command in this
script\footnote{https://github.com/haotian-liu/LLaVA/} with minor modification, we will elaborate on this modification in section \ref{fine-tuning}'s fine-tuning practice.

The model update process is an iterative cycle that repeats periodically, ensuring that the robot remains accurate in a dynamic environment.

    \section{Experiments}
    For the datasets, we collect 10 similar medicine bottles, inlcuding Vitamin B1 (Thiamine), Vitamin B6 (Riboflavin), \textit{etc.}, with each bottle accompanied by its name, color and usage. These medicine bottles, as ordinary people would not intentionally pay attention to, represent a highly challenging novel scenario for the robot, as it is highly likely that it has never encountered these samples in the pre-training dataset.
    
    Our hardware setting contains a camera, a speaker, 
    a microphone and also a laptop to connect the server running LLaVA-NeXT model. 
    We conducted three rounds of experiments to assess our approach's effectiveness. Participants will record two scores based on each dialogue experience:1).score of whether the robot generated the correct answer, which is simply right or wrong, which will be calculated for the overall accuracy;2).score of the dialogue process, which is on a scale of 10, a higher score indicates better accuracy and user experience.

\begin{figure}[htbp]
    \centering
    \includegraphics[width=0.8\linewidth]{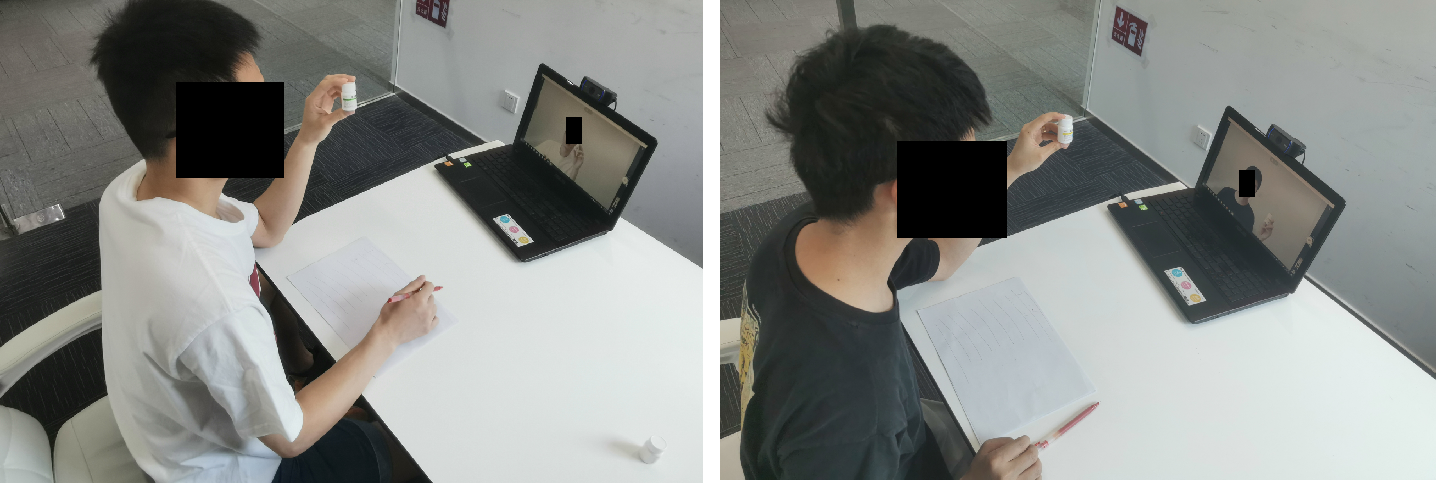} % 
    \caption{A typical scenario of our experiment.
    }
    \label{scenario}
\end{figure}

    Specifically, in each round of testing, we invited 25 individuals. During each dialogue, the participants would hold one medicine bottle 
    at a time and engage in a conversation with the robot, asking for the name, color and uses of the medicine they were holding. 
    Participants can hold the medicine bottle in any posture and pose questions about these three attributes(name, color and uses)
    to the robot in any form. They would record 1) whether the robot generated the correct answer, and 2) score the dialogue process, the score is based on the overall dialogue experience and the final answer given by the robot. Each participant would conduct such dialogues with all 10 medicine bottles. A typical scenario of our experiment is shown in figure~\ref{scenario}.

\begin{figure}[t!]
\centering
\includegraphics[width=1\linewidth]{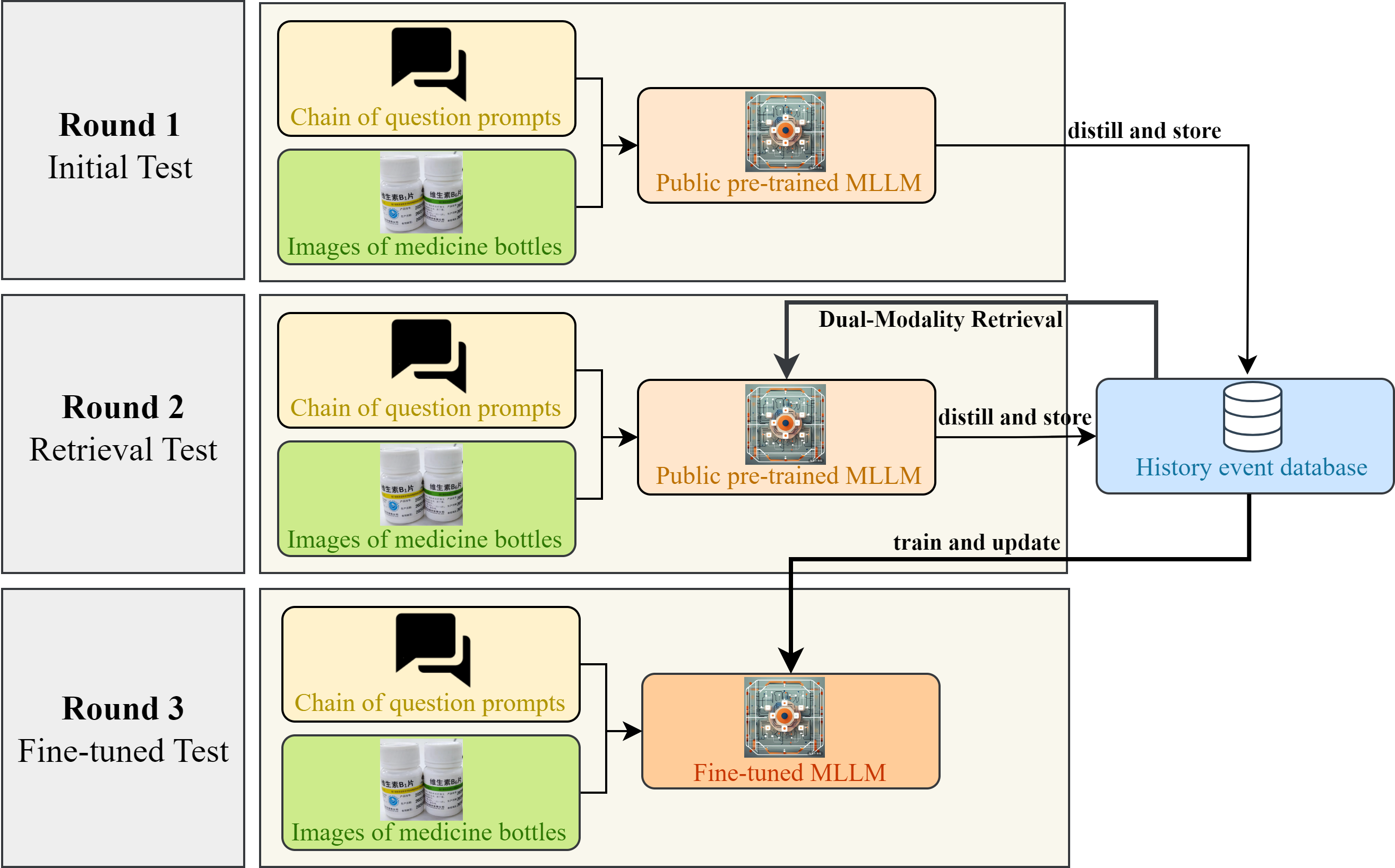} % 
\caption{An illustration of our experiment flow. }
\label{test-flow}

\end{figure}

An illustration is shown in figure~\ref{test-flow}. In the first round, the robot is restricted from retrieving any past interaction events, it just utilizes an ordinary
public pre-trained MLLM equipped with our chain of question prompts; In the second round, the robot simply refers to these interaction events collected during the first round while without fine-tuning on these events;
In the third round, the robot utilizes a MLLM fine-tuned on the events collected from the first two rounds, 
and also, it is equipped with our chain of question prompts, without retrieving any past interaction events.

\textbf{Round 1: Initial Test.} In this round, the robot, utilizing the LLaVA-NeXT,
has neither seen our collected medicine bottles 
nor been specifically fine-tuned on it. 
The participants' purpose is to ask the robot the name, 
color and uses of the medicine bottle they are holding and to score the dialogue process and the robot's final answer. The score ranges from 1 to 10, with higher scores representing greater satisfaction. The participants will also record whether the robot successfully generated the correct answer to the question.

\textbf{Round 2: Retrieval Test.} In this round, after the first round of experiment, a complete database
of history events is available for retrieval.
The 25 participants, each assigned 10 medicine bottles,
will still ask the robot the name, color and uses of the medicine they are holding, record the answer and score the response and the dialogue process. Similarly, this round of dialogue will also be distilled into the corresponding format and 
stored in the history event database.
\begin{figure}[t!]
    \centering
    \includegraphics[width=1\linewidth]{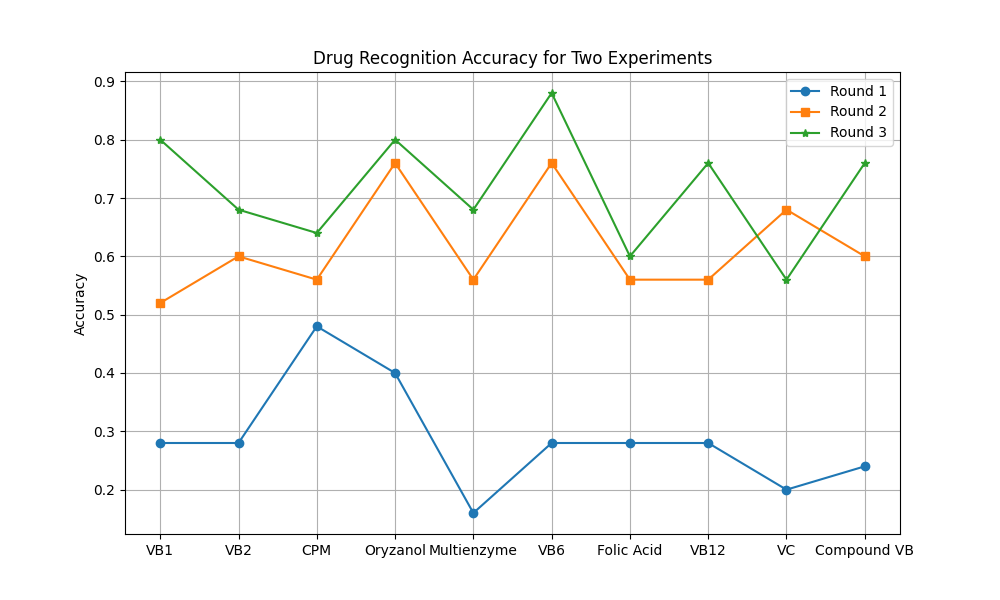} % 
    \caption{Accuracy. VB for Vitamin B, CPM for Chlorpheniramine Maleate. 
    The average accuracy of all 10 medicines in each round are 0.288, 0.616, 0.716
    , higher accuracy indicates better performance. 
    }
    \label{0or1}
    
\end{figure}

\textbf{Round 3: Fine-tuned Test.} After the first two rounds of dialogue, the data, stored in the history event database, will be used for fine-tuning the robot, finetuning details will be discussed in section \ref{fine-tuning}'s fine-tuning practice. 
Once the training is complete, the model's weights will be updated, and then the current round of testing experiments will begin.
Similarly, this round the robot is set as the first round: it will not retrieve any past interaction events, and is solely equipped with chain of question prompts. The participants, again with the 10 bottles, will ask questions, record the answers, and score the dialogue process and the response. 

\subsection{Results}
\textbf{Accuracy of All 3 Rounds.}% After all the 3 rounds of tests, 
We collected all the records of whether the robot correctly answered questions about the medications and calculated the accuracy for each round, as shown in figure~\ref{0or1}.

In the first round, the robot had almost never encountered these rare categories of
medications, and it was challenging for it to perceive these medication bottles well with its intrinsic knowledge, 
resulting in very low accuracy. In the second round, as there was already a database 
of past interaction events from the first round available for retrieval, the robot, even without training on this data,
was able to significantly improve its performance by leveraging dual-modality retrieval. In the third round, due to our 
proprietary history event database collection and model fine-tuning, the robot was able to achieve the best results solely based on its own intrinsic knowledge just learned, without relying on any past event databases during user interactions.

\begin{table}[htbp]
    \centering
    \caption{Tables of results. Better results within each setting are bolded.}
    \label{tab:main_table}
    \begin{subtable}{1\linewidth}
    \centering
    \caption{Average and Variance of all 3 rounds' scores.% The metrics are on a scale of 10, a higher average score indicates better accuracy and user experience, while a lower variance score for better stability.
    }
    \label{average-variance}
    \begin{tabular}{l l l l}
        \toprule
        & Round 1 & Round 2  &Round 3 \\
        \midrule
        average& 3.34  &5.38 &\textbf{7.1} \\
        variance  & 4.90&8.40 &\textbf{2.89}  \\
        \bottomrule
    \end{tabular}
    \end{subtable}\hfill
    \begin{subtable}{1\linewidth}
    \centering
    \caption{Ablation study of different fine-tuning strategies.
    % frozen denotes the LLaVa-NeXT trained following the official fine-tuning strategy, while updatable for not freezing any weights. 
    }
    \label{ablation}
    \begin{tabular}{p{1.3cm} p{1.5cm}p{1.5cm}p{1.25cm} }
        \toprule
        % \hline
        visual   encoder& average accuracy &average score &variance scores \\
        \midrule
        frozen & 0.658 &6.66 & 3.824\\
        updatable  & \textbf{0.716}&\textbf{7.1} &\textbf{2.89}  \\
        \bottomrule
    \end{tabular}
    \end{subtable}
    \end{table}

\textbf{Scores of All 3 Rounds.}% After all the 3 rounds of tests, 
We collected all the participants' scoring results from all three rounds of 
testing and calculated the average and variance of 
the scores for each round. The average serves as an intuitive measure of the model's accuracy 
and user experience during the 
dialogue process. Additionally, the magnitude of the variance can also indicate the stability of the robot's capability: a larger 
variance suggests that the model's performance in answering questions fluctuates more, which in turn means that it cannot
guarantee a good user experience.
The results are shown in Table \ref{average-variance}.

As can be seen from the calculated results, In the first round, the robot got the lowest average score 
and middle variance, which indicates that it 
performs consistently poorly. In the second round, with the assistance of past interaction events, the robot got a higher average score than round 1,
but 
got worse variance, which indicates that the history event database helps a lot in improving the model's performance,
but at the same time, the model 
fails to fully utilize these interaction events, leading to significant performance fluctuations across different dialogues. As for the last round, the model
got the highest average score and the lowest variance score, which proves that the fine-tuning training process not only enables
the model to fully utilize these interaction events but also ensures that it can consistently and stably generate correct answers to provide a better
user experience.

\label{fine-tuning}
\textbf{Fine-tuning Practice.} We utilized LLaVa-NeXT's official codes to fine-tune the models leveraging the collected 
interaction events. However, It's worth noting that its end-to-end fine-tuning strategy always keeps the visual encoder weights frozen, and continues to update both the pre-trained weights of the projection layer and LLM~\cite{llava}, we believe that there may still be room for improvement with this approach for tasks involving novel scenarios, because the public LLaVa-NeXT is pre-trained on the general corpus.

We conducted an ablation study
on our collected interaction events: After the former two rounds' test, we collected all the past interaction events to fine-tune the model.
On one hand, we followed the LLaVa-NeXT's original fine-tuning approach, freezing the weights of the visual encoder
and fine-tuning weights of the projector layer and the LLM. On the other hand, we fine-tuned an additional model without freezing any weights, 
meaning that the visual encoder, the projector layer, and the LLM were all involved in training and parameter updates. 
Then, we conducted the experiment round 3
using these two fine-tuned models respectively and inferred which approach was better based on the users' scores.
The results are shown in Table~\ref{ablation}.

As can be seen from the results, the LLaVa-NeXT fine-tuned without freezing any weights is superior in every aspect.
% This result is, to some extent, predictable. 
Since our dataset consists of highly novel scenarios, 
especially with pixel differences 
between different objects being very minor, traditional fine-tuning, 
which only updates the weights of the projector 
and the LLM, struggles to effectively distinguish the differences between different objects. 
In contrast, updating the 
weights of the visual encoder simultaneously can help to
extract more informative features for perception and recognition.

\section{Conclusion}
% In this paper, 
We proposed an interactive learning-based robot system powered by a MLLM for enhancing performance 
and adaptability from everyday dialogues. Our system accumulates interaction events, distills and annotates them based on user feedback.
The system can generate answers based on past interaction events in diverse environments
to avoid repeating the same mistakes corrected before.%, thereby offering a superior approach for robots to adapt and providing users with better experience before the model is updated. 
Furthermore, model update will be initiated when the number of accumulated interaction events reaches a threshold.
% , ensuring ongoing optimization of model performance. 
Our experiments also validate the effectiveness of our approach. We hope that our method can make a tiny contribution to this field,% thereby 
inspiring further research and
development\footnote{The authors have no competing interests to declare that are relevant to the content of this article.}.

\bibliographystyle{splncs04}
\bibliography{custom}

\begin{thebibliography}{10}
\providecommand{\url}[1]{\texttt{#1}}
\providecommand{\urlprefix}{URL }
\providecommand{\doi}[1]{https://doi.org/#1}

\bibitem{justAsk}
Chi, T.C., Shen, M., Eric, M., Kim, S., Hakkani-Tur, D.: Just ask: An interactive learning framework for vision and language navigation. In: Proceedings of the AAAI conference on artificial intelligence. vol.~34, pp. 2459--2466 (2020)

\bibitem{mqa}
Deng, Y., Guo, D., Guo, X., Zhang, N., Liu, H., Sun, F.: Mqa: Answering the question via robotic manipulation. arXiv preprint arXiv:2003.04641  (2020)

\bibitem{driess2023palm}
Driess, D., Xia, F., Sajjadi, M.S., Lynch, C., Chowdhery, A., Ichter, B., Wahid, A., Tompson, J., Vuong, Q., Yu, T., et~al.: Palm-e: An embodied multimodal language model. arXiv preprint arXiv:2303.03378  (2023)

\bibitem{lora}
Hu, E.J., Shen, Y., Wallis, P., Allen-Zhu, Z., Li, Y., Wang, S., Wang, L., Chen, W.: Lora: Low-rank adaptation of large language models. arXiv preprint arXiv:2106.09685  (2021)

\bibitem{kang2020natural}
Kang, Y., Cai, Z., Tan, C.W., Huang, Q., Liu, H.: Natural language processing (nlp) in management research: A literature review. Journal of Management Analytics  \textbf{7}(2),  139--172 (2020)

\bibitem{robotvqa}
Kenfack, F.K., Siddiky, F.A., Balint-Benczedi, F., Beetz, M.: Robotvqa—a scene-graph-and deep-learning-based visual question answering system for robot manipulation. In: 2020 IEEE/RSJ International Conference on Intelligent Robots and Systems (IROS). pp. 9667--9674. IEEE (2020)

\bibitem{llava-next}
Liu, H., Li, C., Li, Y., Li, B., Zhang, Y., Shen, S., Lee, Y.J.: Llava-next: Improved reasoning, ocr, and world knowledge  (2024)

\bibitem{llava}
Liu, H., Li, C., Wu, Q., Lee, Y.J.: Visual instruction tuning. Advances in neural information processing systems  \textbf{36},  2434--2447 (2024)

\bibitem{MuModaR}
Menezes, J.C.: Mumodar: Multi-modal framework for human-robot collaboration in cyber-physical systems. In: Companion of the 2024 ACM/IEEE International Conference on Human-Robot Interaction. pp. 755--759 (2024)

\bibitem{multimodal-robot1}
M{\"u}ller, S., Wengefeld, T., Trinh, T.Q., Aganian, D., Eisenbach, M., Gross, H.M.: A multi-modal person perception framework for socially interactive mobile service robots. Sensors  \textbf{20}(3), ~722 (2020)

\bibitem{paranjape2023art}
Paranjape, B., Lundberg, S., Singh, S., Hajishirzi, H., Zettlemoyer, L., Ribeiro, M.T.: Art: Automatic multi-step reasoning and tool-use for large language models. arXiv preprint arXiv:2303.09014  (2023)

\bibitem{clip}
Radford, A., Kim, J.W., Hallacy, C., Ramesh, A., Goh, G., Agarwal, S., Sastry, G., Askell, A., Mishkin, P., Clark, J., et~al.: Learning transferable visual models from natural language supervision. In: International conference on machine learning. pp. 8748--8763. PMLR (2021)

\bibitem{Improving-grounded}
Thomason, J., Padmakumar, A., Sinapov, J., Walker, N., Jiang, Y., Yedidsion, H., Hart, J., Stone, P., Mooney, R.J.: Improving grounded natural language understanding through human-robot dialog. In: 2019 International Conference on Robotics and Automation (ICRA). pp. 6934--6941. IEEE (2019)

\bibitem{mllms}
Wu, J., Gan, W., Chen, Z., Wan, S., Philip, S.Y.: Multimodal large language models: A survey. In: 2023 IEEE International Conference on Big Data (BigData). pp. 2247--2256. IEEE (2023)

\bibitem{yang2023mm}
Yang, Z., Li, L., Wang, J., Lin, K., Azarnasab, E., Ahmed, F., Liu, Z., Liu, C., Zeng, M., Wang, L.: Mm-react: Prompting chatgpt for multimodal reasoning and action. arXiv preprint arXiv:2303.11381  (2023)

\bibitem{llm-survey}
Yao, Y., Duan, J., Xu, K., Cai, Y., Sun, Z., Zhang, Y.: A survey on large language model (llm) security and privacy: The good, the bad, and the ugly. High-Confidence Computing p. 100211 (2024)

\bibitem{distillingRobot}
Zha, L., Cui, Y., Lin, L.H., Kwon, M., Arenas, M.G., Zeng, A., Xia, F., Sadigh, D.: Distilling and retrieving generalizable knowledge for robot manipulation via language corrections. arXiv preprint arXiv:2311.10678  (2023)

\bibitem{vison-language-navigation}
Zhu, F., Zhu, Y., Chang, X., Liang, X.: Vision-language navigation with self-supervised auxiliary reasoning tasks. In: Proceedings of the IEEE/CVF conference on computer vision and pattern recognition. pp. 10012--10022 (2020)

\end{thebibliography}

\end{document}